\documentclass[aps,rmp,twocolumn]{revtex4}

\include{graphicx}
\begin{document}


\title{Quantum Erasure:  Quantum Interference Revisited\footnote{Unedited and unillustrated version of ``Quantum Erasure", \textit{American Scientist} \textbf{91} 336-343 (2003).}}

\author{Stephen P. Walborn}
\email[]{swalborn@fisica.ufmg.br}
\affiliation{Universidade Federal de Minas Gerais, Caixa Postal 702, Belo Horizonte, MG
30123-970, Brazil}
\author{Marcelo O. Terra Cunha}
\affiliation{Universidade Federal de Minas Gerais, Caixa Postal 702, Belo Horizonte, MG
30123-970, Brazil}

\author{Sebasti\~ao  P\'adua}
\affiliation{Universidade Federal de Minas Gerais, Caixa Postal 702, Belo Horizonte, MG
30123-970, Brazil}
\author{Carlos H. Monken}
\affiliation{Universidade Federal de Minas Gerais, Caixa Postal 702, Belo Horizonte, MG
30123-970, Brazil}

\date{ April 20, 2003}

\begin{abstract}
Recent experiments in quantum optics have shed light on the foundations of quantum physics.  Quantum erasers - modified quantum interference experiments - show that quantum entanglement is responsible for the complementarity principle.   
\end{abstract}


\maketitle

It may be somewhat surprising that Thomas Young's double-slit experiment - a staple in the freshman physics laboratory - would be such an invaluable testing ground for the foundations of quantum physics.  Yet the quantum version of the double-slit experiment has been at the center of many debates over the fundamentals of quantum physics since the theory was born, nearly a century ago.  In fact, Young's experiment embodies the very nature of quantum physics.  Last year, the readers of Physics World magazine voted Young's double-slit experiment with electrons ``the most beautiful experiment" in physics.  The significance of Young's experiment lies in the fact that interference is a phenomenon exhibited only by waves.  The puzzle that quantum physics presents is that a particle, which is usually thought of as an indivisible, localized object, can also behave like a classical wave, which interferes and diffracts.   In ``the most beautiful experiment", electrons pass through the slits like waves and are detected like particles!  This interference behavior is perhaps the greatest mystery in quantum theory.  In fact, Nobel Prize-winning physicist Richard Feynman has called quantum interference ``the only mystery" in quantum physics.   Recently, some progress has been made in the understanding of these interference effects within the foundations of quantum theory.  Experiments called quantum erasers - modified versions of Young's experiment - have shed light on the foundations of quantum physics.  However, before we explain the notion of quantum erasure, we take a detour to explore the concept and the history of classical and quantum interference.               
\par
In the freshman physics laboratory, the double-slit experiment is quite simple.  A laser beam is directed onto two closely spaced transparent slits that are etched into an opaque microfilm.  The slits and their spacing are about a tenth of a millimeter wide.  The laser beam is scattered by this ``double-slit" and a pattern of alternating bright and dark stripes - commonly called interference \emph{fringes} - is projected onto a distant viewing screen.  Understanding the reason for this interference is not difficult:  the paths from each slit to a given observation point are not necessarily equal, so light beams traveling from each slit arrive with different phases of propagation.  These light beams interfere depending upon the difference in their phases: either constructively, resulting in an interference maximum (bright stripe) or destructively, resulting in an interference minimum (dark stripe).  Even if you have never set foot in a physics laboratory, you have undoubtedly observed interference.  Interference effects cause many common optical phenomena, such as the color patterns seen in soap bubbles or in the oily puddles in the parking lot of a gas station. 
\par
  Another way to visualize interference is to imagine a water wave incident on a wall with two vertical openings.  When the wave front encounters the wall, a part of the wave goes through each opening, while the rest is reflected.   The two sections that pass through the slots will meet up again a distance later and combine, or interfere.  If a classical particle, say a tiny dust particle or even a tennis ball, is launched at the wall, it will either go through one of the openings or bounce back.  To interfere, the particle would have to ``pass through both slits at the same timeÓ!  So it is very surprising and almost unbelievable that when this particle is instead an electron, for example, it interferes like a wave. 
  \par          
A fundamental result of quantum theory is that light is made up of tiny quanta of energy Ð ``particles of light" - called photons. In 1909, Geoffrey Taylor demonstrated diffraction of individual photons using the tip of a needle.  Diffraction occurs when a wave passes through a tiny aperture or object.  The diffraction pattern is similar to an interference pattern:  maxima and minima are due to the interference of different parts of the transmitted wave that meet at the detection screen.  
  \par
  What happens when Young's experiment is repeated using individual photons instead of an intense light beam?  An attenuated light source ensures that only one photon is incident on the double slit at a time.  After recording data for many photons, the resulting pattern of individual points (each corresponding to the detection of one photon) on the photosensitive screen is identical to that of an intense light beam, interference fringes and all.  This seems to imply that the individual photons had ``passed through both slits at the same time" and ``interfered with themselves", a seemingly astounding feat, even for something as aloof and mysterious as the photon.  To date, variations of the quantum double-slit experiment have been performed using many different types of particles, including photons, electrons, neutrons and even large carbon-60 fullerenes.  All results confirm the counter-intuitive result that, at the quantum level, particles ``interfere with themselves" just like classical waves.  
\section{The quantum coin toss}
To further understand why the interference of quantum particles is an unexpected result, here is a simple example.  Consider the usual coin toss, of the sort that takes place at the start of an NFL football game, where the coin has the same chance of giving heads or tails.  The probability (call it $P$(heads)) that a coin lands heads is thus 50 \%.  Likewise, the probability that the coin gives tails is $P$(tails) = 50\%.  Obviously, there are only two possible outcomes, the coin must land either heads or tails, so the total probability to give heads or tails is just the sum of the individual probabilities: $P$(heads or tails) = $P$(heads) + $P$(tails) = 100\%.  The quantum double-slit is a type of ``quantum coin toss", and so we can make a similar analysis.  Given a certain position on the detection screen, one can try to assign a probability $P$(slit 1) or $P$(slit 2) that a photon detected at that point on the screen passed through slit 1 or slit 2.  Here comes the surprising result:  unlike the coin toss, the total probability to register a photon is not equal to the sum of the individual probabilities: $P$(slit 1 or slit 2) $\neq$ $P$(slit 1) + $P$(slit 2).  
  \par
  The physical principle responsible for this strange behavior is called \emph{superposition}, which says that wavelike events combine according to their probability amplitudes, not their probabilities.  Let's denote the probability amplitude with the letter $A$.  The probability amplitude for a photon to pass through slit 1 is $A$(slit 1) and $A$(slit 2) is the amplitude for the photon to pass through slit 2.  One difference between a probability and a probability amplitude is that the amplitudes are now complex numbers, to incorporate the concept of phase.  The total probability amplitude for a photon to pass through slit 1 or slit 2 is $A$(slit 1 or slit 2) = $A$(slit 1) + $A$(slit 2).  The probability for a given event is then obtained by calculating the ``absolute square" of the corresponding probability amplitude: $P = |A|^2$.  Thus, the total probability to detect a photon is $P$(upper or lower) =$ | A$(slit 1) + $A$(slit 2)$|^2$.  Computing this probability gives rise to quantities, not present in the NFL coin toss example above, which are responsible for the interference effects.  Quantum particles - electrons, photons, etc - interfere because they behave according to the superposition principle, which describes the physical phenomenon of waves.  Thus, when you flip a  ``quantum coin", it can give both heads and tails at the same time.  
  \section{Particles or waves?}
By the time Young performed his experiment in 1801, physicists had been debating the nature of light for many years.  The question was:  Is light made of waves or particles?  Some scientists, such as Isaac Newton (1717), believed light was made up of tiny classical particles, like particles of dust.  The movement of each particle traced out a trajectory, called a ray.   Others, such as Dutch physicist Christian Huygens (1690), advocated a classical wave theory, like water waves or oscillations on a stretched string.  Each theory was able to explain some of the phenomena observed up until that time, such as shadows, refraction and reflection. But when Thomas Young showed that a beam of light interferes with itself, which a classical particle could never do, the particle theory was laid to rest.  That is, until Albert Einstein came along.  
 \par
At the end of the nineteenth century, German physicist Max Planck was concerned with the following problem:  explain the color spectrum of radiation emitted by a ``blackbody".  A black body is basically a metal box kept at a certain temperature with a small hole allowing radiation to escape.  Planck was interested in the color spectrum emitted by the box with respect to its temperature.  Using classical radiation theory to describe blackbody radiation gave an inaccurate result known as the ultra-violet catastrophe. To accurately explain the radiation spectrum, Planck proposed the idea that light is made up of discrete energy units, or quanta, which we now call photons.  Planck was reluctant to accept his own idea, which he thought of as a mathematical``trickÓ which happened to fit the experimental data. Planck tried vigorously to explain blackbody radiation using other physical concepts.  
Shortly thereafter in 1905, Albert Einstein, in addition to publishing his seminal works on relativity and Brownian motion, applied Planck's revolutionary idea to explain the photoelectric effect, the work for which he was later granted the Nobel prize in 1921 (Planck had won the Nobel prize for his research 3 years earlier).  Though Planck was the first to propose the idea of quanta, it was Einstein who embraced the idea, and his work along with Planck's forced the physics community to accept it.  It was the dawn of quantum physics.  
\section{Matter waves matter}
 \par
Photons and other quantum particles are absorbed in discrete units of energy.  The detection of a particle corresponds to a tiny point on some type of detection screen.  But above we stated that quantum particles interfere with themselves just like waves.  How can quantum objects have both particle and wave characteristics?   In other words, how can a photon interfere with itself when passing through a double-slit but later appear as a tiny point on a photosensitive film?  This paradox is known as wave-particle duality, and is one of the cornerstones of quantum theory.  Wave-particle duality is often revealed through another underlying concept called the \emph{complementarity principle}.   
 \par
In quantum physics, physically measurable quantities (such as position, momentum, etc.) are often called \emph{observables}.  The complementarity principle states that the more we know about a given observable, the less we know about its complement.  For example, if we measure the exact position of an object at an instance in time, then we can have no knowledge of the object's momentum at that instance.  Position and momentum are called complementary observables.  To avoid any confusion with the classical and quantum aspects of the word``particle", we have now resorted to using the word``object" to describe a quantum particle - meaning a photon, an electron, a neutron, etc.  
 \par
The concept of position corresponds to a point in space.  Imagining again a water wave, or a wave on a stretched string, with series of peaks and troughs, it is easy to see that a wave does not have a well-defined position in this sense.  A wave, such as those that can be seen crashing onto a sandy beach, can be localized to within a certain region, but not to a point.  A classical particle does possess a well-defined position, and using the laws of classical physics, one can calculate the particle's trajectory and know its position at all instances in time.  Therefore, position is identified as a particle-like property.    A wave, on the other hand, can be described in terms of its frequency, wavelength, amplitude and phase.  In 1927 Louis de Broglie characterized the wavelength (now known as the de Broglie wavelength) of a quantum object with its momentum, work for which he was later granted the Nobel Prize.  Consequently, in quantum physics, momentum is a wave-like property.  Hence the complementarity of position and momentum leads to wave-particle duality:  quantum objects can behave as either particles or waves.  The observed behavior depends on what type of measurement the experimenter chooses to make:  if a particle-like property such as position is measured, then the quantum object behaves like a particle.  Likewise if we choose to observe a wave-like property, such as momentum, the observed behavior is wave-like.  Moreover, quantum physics does not provide us with the means to make any definite statement about the properties of the quantum object before we measure it.  The observation of a wave-like property does not imply that the quantum object was behaving as a wave just before the measurement.    
 \par
If this all sounds pretty unbelievable to you then you are in good company.  Many of the founding fathers of quantum theory were not very satisfied with this state of affairs either, including Einstein, whose intellectual battles with Danish physicist Niels Bohr are the stuff that many physics books are made of.  Bohr was the greatest proponent of the idea of complementarity, an idea that Einstein was reluctant to accept.  Einstein could not come to terms with the idea that what we observe and consequently call``reality" seems to be  based solely on the manner in which we choose to look.  Moreover, he was bothered by the fact that according to quantum theory, this reality only exists while we are observing.  He expressed his discontent to Abraham Pais by asking: ``Do you believe that the moon exists only when you look at it?"  Einstein did not accept that quantum theory was a complete description of nature.  Interestingly, it was Einstein's dissatisfaction that motivated and still motivates much of the modern research in quantum mechanics.      
 \par      
Many of the great Einstein-Bohr dialogs took place at the Solvay conferences in the 1920's.  On several occasions, Einstein thought he could poke holes in Bohr's so-called Copenhagen Interpretation of quantum theory.  Throughout the history of physics, much of the discussion and debate over the nature of the world is done through examples and counter-examples of \emph{gedanken} experiments:  idealized thought experiments.  One of Einstein's famous examples is the following.  Repeat the quantum version of Young's experiment, but this time the double slit is suspended by sensitive springs so that it is free to move back and forth.  An incident photon, scattered by the slits, suffers a change in momentum, which is absorbed by the double slit apparatus, giving it a slight kick.  One could then measure the recoil of the slit apparatus together with the photon's position on the detection screen and infer the photon's trajectory, a particle-like property.  The trajectory of the photon itself should not be altered by this measurement, so the interference fringes - a wave-like property - should still be observed.  From the spacing between the interference fringes one can calculate the (de Broglie) wavelength and thus the momentum of the photon.  In such a way it should be possible to observe the characteristic interference fringes and calculate the momentum as well as know the photon's trajectory.  The complementarity principle must be a hoax! 
 \par
  Bohr later pointed out, however, that Heisenberg's uncertainty relation prevented one from seeing interference fringes and determining the photon's trajectory simultaneously.  The uncertainty relation is a quantitative statement about the best precision with which one can measure complementary observables.  The recoil of the double-slit apparatus (an indicator of the momentum of the photon) disturbs the system creating an uncertainty in the detection of the photon's position on the detection screen.  This uncertainty is great enough to ``wash out" or blur the interference fringes to such a degree that they no longer appear.  Any attempt to measure the photon's trajectory disturbs the system and prevents the observation of interference fringes.  All ideas similar to that of Einstein's have failed due to similar arguments.  For many years it was thought that the uncertainty relation was the mechanism responsible for the complementarity principle.  The question remained:  are we able to mark the particle's path (1) without altering it's trajectory and (2) in such a way that we can get around the uncertainty principle?                       
\section{Quantum Erasure}
Roughly twenty years ago, physicists Marlan O. Scully and Kai Dr\"uhl (at the Max-Planck Institut f\"ur Quantenoptik and University of New Mexico)  shook the physics community and strengthened the foundations of quantum physics, when they introduced the idea of quantum erasure.  The logic of quantum erasure is the following:  if the information providing the object's trajectory can be determined without significantly perturbing it, then the interference disappears, but the``erasure" of this information should bring the interference back.  Through the introduction of this new concept, they showed that the complementarity principle plays a much more fundamental role in quantum physics than the uncertainty relation.  
  \par
Later, Scully, with Berthold-Georg Englert and Herbert Walther (both at the Max-Planck Institut f\"ur Quantenoptik) proposed a way to bring this about using Rydberg atoms as the interfering objects.  Rydberg atoms are excited at very high electron energy levels (for example $n=50$) with long decay times. The atoms are incident on a double-slit.  Two microwave cavities, made of a pair of microwave high reflectors, are then placed one behind each slit.  The microwave cavities serve as path markers.  When an atom passes through a cavity it emits a photon, which remains in the cavity.  In this process, the atomÕs trajectory is not disturbed.  By simply looking to see which cavity contains the photon, it would be possible to know where the atom has been.  So far the Scully-Englert-Walther experiment has never been realized in the laboratory.    
However, we have succeeded in performing an experiment that is analogous to their proposal and much easier to implement experimentally.  However, first we must digress briefly to explain the concept of polarization.  
  \par
The electromagnetic field, that is light, as well as the photon, has an internal  property called polarization.  In classical optics, light is viewed as a transverse electromagnetic wave and polarization refers to the direction in which it oscillates.  A field that oscillates in a specific manner is said to be polarized.  A field with linear polarization oscillates back and forth along a certain direction, perpendicular to the propagation direction, while a field with circular polarization oscillates in a circular pattern.  Right-circular polarized light oscillates in the clockwise direction, while left-circular polarized light oscillates in the counter-clockwise direction.  A circular polarized light beam can be described as a superposition of horizontally and vertically polarized beams that are a quarter cycle (or quarter wavelength) out of phase with each other.  For right-circular polarization the vertical component is a quarter cycle ahead of the horizontal component, while for left-circular polarization the vertical component is a quarter cycle behind the horizontal component.  Other commonly used polarization directions are the diagonal directions, $45^\circ$ and $-45^\circ$. The diagonal directions are superpositions of horizontal and vertical components just like the right- and left-circular polarizations, only now the horizontal and vertical components are in phase ($45^\circ$) or one-half cycle out of phase ($-45^\circ$) with each other.    
Optical components called wave plates are used to change the polarization, while the propagation direction of the electromagnetic field is left untouched. A quarter-wave plate can be used to convert a linearly polarized beam into a circularly polarized beam.  Another commonly used optical components is a polarizer, which acts as polarization filter, allowing only light with a given polarization to pass.  For example, if a circularly polarized beam is directed onto a horizontal polarizer, the beam which exits is horizontally polarized and half as intense as the input beam.  Polarizing sunglasses use this concept to eliminate glare from reflective surfaces. 
\par
Now imagine that we repeat Young's experiment with photons polarized linearly in the vertical direction, and we observe interference fringes on a distant screen.  Suppose now that we insert two quarter-wave plates, one behind each slit, in such a way that plate 1 transforms the vertically polarized photons into right-circularly polarized photons, while plate 2 transforms the vertically polarized photons into left-circularly polarized photons.  The result is that no interference pattern is observed at the detection screen.  Instead, after many photons, we will observe a distribution of photon detections that produces the famous bell-shaped curve.  The pattern looks something like a mountain peak, with a maximum in the middle, where photons from each slit will hit.  There is only one peak because the two slits are very close together.  If the slits were well separated, two peaks would appear.  
  \par
  What happened to the interference?  The quarter wave plates have marked the polarization of the photons.  All we have to do is measure the circular-polarization direction (left or right) of the photons at the screen and we will know through which slit the photons have passed.  Since right- and left-circular polarizations oscillate in opposite directions, they are completely distinguishable from each other.  Moreover, the quarter-wave plates do not alter the propagation direction of the photons.  It is important to note that we don't actually have to measure the polarization direction in order to destroy the interference pattern.  It is enough that the so-called which-path information is available to us.  Playing dumb will not restore the interference fringes.  
\par
  One might note that this experiment could just as well have been performed using an intense classical light beam.  We have chosen to use quantum interference - photons - because the question as to which slit the beam of light has passed through has no significance in classical optics, where a beam of light is always a wave, and thus the concept of position is meaningless.      
\subsection{Interference is Ignorance}
  \par
What happens if we instead measure polarization in the horizontal direction?   If we limit our observation apparatus to only horizontally polarized photons, then we will again see interference fringes.   But how can that be?  The quarter-wave plates have marked the photons path.  Simply ignoring the information does not bring back interference.  Why do we observe interference if we measure horizontal polarization?  
  \par
  Both right- and left-circular polarizations have a horizontal component and thus observation of a horizontally polarized photon tells us nothing about through which slit the photon has passed.   The key here is that measuring horizontal polarization erases the which-path information (hence the name``quantum erasure").    If we tried to measure right- or left-circular polarization again after the horizontal polarizer, we would gain nothing in the way of which-path information.         
\par
  Similarly, if we choose to measure vertical polarization, we again erase the which-path information and restore interference.  However, in this case we observe interference in the form of \emph{antifringes} that are completely out of phase with those observed with horizontal polarization, meaning that where we had observed an interference maximum (a bright spot) we now observe a minimum (a dark spot), and vice versa.    As it so happens, the sum of these interference patterns reproduces the``mountain peak" pattern that one would obtain had no polarization measurement be made.  This is the essence of quantum erasure.  
  \par
  Our choice of polarization measurement divides the experimental results into subsets.  Some of these subsets give interference fringes, as in the case where we measure horizontal or vertical polarization, while other subsets give which-path information, as when we measure either right- or left-circular polarization.  If we add together the measurement results for the cases which give interference, the sum reproduces the mountain peak, as though we had not made any polarization measurement.  Similarly, if we add together the measurement results for the cases which give which-path information, we obtain the same result.  
  \par
We observe interference because the two possibilities corresponding to slit 1 and slit 2 are at least somewhat indistinguishable, that is, our choice of measurement cannot tell us with certainty through which slit a detected photon has passed.  If the two possibilities are completely indistinguishable, as is the case when we measure horizontal or vertical polarization, we observe perfect high-contrast interference fringes.  Likewise interference is completely destroyed when the two possibilities are distinguishable, meaning that our measurement apparatus is capable of telling us with certainty through which slit the photon has passed, as is the case when we measure circular polarization.  There exist quantitative mathematical relationships governing the contrast of interference fringes and amount of which-path information we can observe simultaneously.     
  \par
What prevents us from observing interference and determining the photon's trajectory in the quantum eraser?  Polarization and position are not complementary observables so there is no place for an explanation based on the uncertainty principle.  Moreover, the fact that we can erase the which-path information and observe interference implies that there is no ``disturbance" involved in the measurements.   Yet the fact remains, we are still unable to obtain which-path knowledge and observe interference fringes simultaneously.  It must be that the complementary principle is enforced through some mechanism more fundamental than the uncertainty relation.    
  \par
If it is not the uncertainty relation, then what is responsible for complementarity?  The answer is \emph{quantum entanglement}.  When a photon passes through the double-slit apparatus (just before it passes through the quarter-wave plates), it is in a superposition of position states:  slit 1 + slit 2.  The quarter-wave plates then perform a conditional logic operation on the photon:  if a photon passes through slit 1 then it emerges with right-circular polarization, and if a photon passes through slit 2 then it emerges with left-circular polarization.  The photon's polarization has become entangled with its path.  The result is a more complicated quantum superposition involving two degrees of freedom:  the photon's path and its polarization.
 \par
Entanglement is the name given to this type of quantum correlation, which is much stronger than any classical correlation.  The reason for this is that entanglement correlates the probability amplitudes, while a classical correlation correlates only the probabilities.  To see this, let's return to the NFL coin toss example, however, imagine now that we have two ``magical" coins, correlated such that when flipped they always give opposite results:  one coin gives heads while the other gives tails.  This is a type of classical correlation.  Individually, each coin still lands heads 50\% of the time and tails the other 50\% of the time.  If you flip both coins and then quickly hide one of them, you can always discover the result of the hidden coin simply by looking at the result of the exposed coin.
\par
The difference between this example of classical correlation and quantum entanglement is that the quantum correlation exists even when you look at superpositions of the individual states.  For example, as we will discuss below, it is possible to create two photons that have entangled polarizations.  That is, if one photon is horizontally polarized then the other is vertically polarized.  If we test both photons individually, there is a 50\% chance that we will measure each photon to be either horizontal or vertical, but we will never find that they are polarized in the same direction simultaneously.  One can test this experimentally using horizontal and vertical polarizers.  Up to this point, this seems to be the same as the magical NFL coins.  However, unlike the NFL coins, it is possible to rotate the polarizers $45^\circ$ so that they measure $45^\circ$ and $-45^\circ$ diagonal polarization.  The photons will display the same correlation:  each individual photon has a 50\% chance to be detected $45^\circ$ diagonally polarized and a 50\% chance to be detected $-45^\circ$ polarized, but they are never polarized in the same direction simultaneoulsy.  Moreover, this is true for any mutual rotation of the polarizers.  This is impossible using the magical NFL coins or any other type of classical correlation!  In this sense, quantum entanglement is much stronger than any classical correlation.           
\par    
As an aside, physicists have known about quantum entanglement since the renowned 1935 paper of Albert Einstein, Boris Podolsky and Nathan Rosen.  Shortly thereafter, Austrian physicist Erwin Schr\"odinger coined the name entanglement.  The fathers of quantum theory, including Einstein and Bohr, puzzled over the nature of entanglement just as they did over quantum superpositions.  Since then, scientists have realized that quantum entanglement is a physical resource that can actually be used in the areas of information technology.  In fact, quantum entanglement is the backbone of a new and rapidly flourishing multidisciplinary field called \emph{quantum information}.
\par
Nearly twenty years ago, several physicists toyed with the idea of using two-level quantum systems, such as the polarization of a photon, as ``quantum bits" in a computer.  Since then the same idea has been applied to many problems in cryptography, communications and computer science, and produced some promising results.  For example, the ``strange" laws of quantum physics provide the only form of cryptography that is proven to be secure, certainly interesting to governments sending top secret information or to anyone making a credit card purchase via the internet.  
\par
Returning now to the quantum eraser, the quarter-wave plates have entangled the photon's path with it's polarization.  Since the two possible polarizations, right- and left-circular, are distinguishable (they oscillate in opposite senses), we can measure the polarization and determine the photon's path with certainty.  Entanglement enforces the complementarity principle by coupling the photons path to different polarizations which are completely distinguishable from each other.  Physicists have now come to roadblock similar to that of Einstein and Bohr.  Is it possible to measure the path of the photon without entangling it?  Entanglement is a fundamental player in the quantum theory of measurement.  In a way, entanglement is the act of measurement:  since it associates the photonÕs path (the slit) with its polarization (which we can measure).  Most physicists would probably bet that the answer to this question is no.         
\section{Twin Photons: an entangled story}
Recently, in the Quantum Optics laboratory at the Universidade Federal de Minas Gerais (UFMG), we took this experiment a step further.  We created a pair of entangled photons using a non-linear optical process called spontaneous parametric down-conversion.  In our experiment, we directed an ultraviolet argon laser beam onto a thin non-linear crystal, which creates two lower energy ``twin" photons.  The two photons, which we will call $a$ and $b$, were generated in such a way that when photon $a$ is found to have horizontal polarization, then photon $b$ will necessarily be vertically polarized.  Likewise, if $a$ is found to have vertical polarization, then $b$ has horizontal polarization.  As discussed above, similar correlations exist for any type of polarization measurements made on the two photons, as long the polarizers measure perpendicular polarizations (horizontal and vertical, left- and right-circular, etc).  These photons are said to be polarization-entangled.  Furthermore, now that the entangled systems are two independent photons, they can be separated any arbitrary distance.  It has been shown experimentally that entangled photons can remain entangled over great distances - the current record, held by physicists at the University of Geneva, is between the cities of Bellevue and Bernex, a distance of about 11 kilometers!  
\par
After creating the entangled photons, we manuevered photon $a$ to the double-slit apparatus (double slit and quarter-wave plates) and then to a photodetector, while photon $b$ passes directly to a separate polarizer and detector.  When the quarter-wave plates were removed, after many photon pairs we observed the usual interference pattern.  However, since we were working with two photons, the photons pairs were detected in coincidence.  Coincidence detection means that we are only interested in the cases where the two photons are registered at their respective detectors simultaneously.  Experimentally, the photons are detected within a small window of time, usually on the order of $10^{-9}$  seconds.  
\par
The biggest experimental hurdle we had to leap was figuring out a way to mount the quarter wave plates in front of the narrow double slit.  To create an observable interference pattern, each slit of the double-slit was about 0.2 millimeters wide, and they were spaced 0.2 millimeters apart.  The usual quarter wave plates that are commercially available are round in shape, about 1 centimeter in diameter and about 2 millimeters thick.  Due their shape and size, it was necessary to modify the wave plates so that they would each cover only one slit.  Using high quality sandpaper, we sanded a straight edge into each wave plate at the required angle, so that they would each cover one slit and join in the narrow space between the slits.  
\par
When we put the quarter-wave plates in place, the interference was destroyed, just like before.  This time, however, the which-path information is available only through coincidence detection.  One quarter wave plate transforms $a$ 's vertical polarization to right-circular, while the other transforms to left-circular.  However, now photon $a$ can be found to be either vertically or horizontally polarized.  For horizontal polarization, the action of the wave plates is reversed.  Thus, measuring only the polarization of photon $a$ will not provide enough information to determine through which slit $a$ has passed.  Through coincidence detection, however, we \emph{are} provided sufficient information. The two-photon logic statements are:  
(1)  ``$a$ right-circular and $b$ horizontal" or ``$a$ left-circular and $b$ vertical" implies that $a$ passed through slit 1 while  (2) ``$a$ left-circular and $b$ horizontal" or ``$a$ right-circular and $b$ vertical" implies that $a$ passed through slit 2.  Interestingly enough, due to the entanglement between $a$ and $b$, we can choose to observe interference or obtain which-path information of photon $a$ based solely on the polarization direction we measure on photon $b$.  Instead of measuring horizontal or vertical polarization of photon $b$, we can measure diagonal (or circular) polarizations, which are superpositions of horizontal and vertical polarizations.  Detecting a diagonally polarized photon erases the which-path information, and consequently we observe interference fringes.  A measurement in the positive diagonal direction ($45^\circ$) gives interference fringes, while a measurement in the negative diagonal direction ($-45^\circ$) gives interference antifringes, exactly out of phase with the fringes.  
\section{Delayed Choice}
Curiously, with this quantum eraser we could actually choose to observe interference or determine photon $a$'s path after photon $a$ has been detected.  Imagine that the detector registering photon $b$ is moved very far away, so that photon$b$ is detected some time after photon $a$.  The experimenter could then wait until after photon $a$ is registered to decide which measurement to perform on photon $b$, and consequently observe interference or determine $a$'s path.   Moreover, we could let photons $a$ and $b$ travel several light minutes away from each other, so that no signal could travel from $a$ to inform $b$ of it's position in the time between $a$ and $b$ are detected.  How can one choose to observe particle-like or wave-like behavior after the interfering particle has already passed through the double-slit?  
When first discussed by American physicist John A. Wheeler in 1978, before the quantum eraser concept was introduced, this type of delayed choice experiment raised serious physical and metaphysical questions.  It seems to imply that the observer could alter photon $a$'s past by choosing how to measure photon $b$.  However, this is not the case.  To explain why, we will tell you a story about the two  most famous people in quantum information:  Alice and Bob.  
\par
Two quantum physicists, Alice and Bob, decide to perform an experiment testing the foundations of quantum mechanics.  Alice sets up a double slit experiment with quarter wave plates, just like we described, in her laboratory on Earth.  Her friend and colleague Bob, who lives on the Mars colony, sends her photons, one by one, across a quantum ``telephone line" that they have set up between their laboratories.  Alice sends the photons, one by one through the double-slit-wave-plate apparatus.  For every photon, she marks it's position, writing something like``Photon 567 landed at position $x  = 4.3$" in her lab notebook.  When Alice later plots her experimental results, she sees that large ``dull" mountain peak, and concludes that there was no interference present in the experiment.  What Bob has not told Alice is that each of her photons is entangled with another photon which Bob has kept for himself.  Bob performs a series of polarization measurements on the photons, about half the time measuring horizontal and vertical polarization and the other half measuring $+45^\circ$ and $-45^\circ$ diagonal polarization.  He records all of his results in his lab book, with statements such as ``Photon 567 ($b$) was detected with horizontal polarization", but he does not inform Alice of his mischief.       
\par
Bob loves magic and a good practical joke.  When visiting Alice one day, she shows him her experimental results on the computer and says ``Look Bob, I performed that quantum eraser experiment and when I plotted my date, all I got was this dull mountain peak, there was no interference".  Bob says "Alice, are you sure" and, after checking his own lab book, he tells her to plot only those photons for which he measured it's entangled partner to be $+45^\circ$ diagonally polarized and ``Ta-Da!" an interference fringe pattern appears.  "Wait Bob, that wasn't there before! How did you make the photons interfere after I already detected them and recorded it all in my lab book?!", Alice exclaims.  Bob, who loves to play for the an audience, replies, "You think that's impressive, well check this out", and he consults his lab book and plots Alice's photons that are paired with photons for which he measured horizontal polarization and ``Ta-Da!" there is no interference pattern, just the smaller (half height) mountain peak.  Alice is perplexed.  Bob, not knowing when to call it quits, does the same for Alice's photons paired with his $-45^\circ$ diagonal polarization measurements and ``Ta-Da!" interference is back, this time in the form of antifringes.  "Bob, that is amazing!  You have control over the past!  While you are at it, can you go back change my lottery ticket from last week to 67-81-138?," Alice asks with a look of awe in her eyes.  Bob is loving the moment, but he is not the greatest magician, and cannot keep his mouth shut about the secret to his tricks.   "No Alice, look, the photons I gave you were actually entangled with photons that I kept for myself.  I did a series of polarization measurements, and recorded my results.  My polarization measurements tell me how to divide up your experimental results so that we can see interference or not, but I cannot change the position at which any photon actually landed," Bob explains.  He shows her by plotting all of the results for which he measured horizontal OR vertical (orthogonal directions) and they observe the large mountain peak.  He then does the same with all results of $+45^\circ$ and $-45^\circ$, and they observe the same mountain peak.  Of course plotting all of the results together regardless of polarization also gives the mountain peak, as Alice had already observed.  So Bob was not able to alter the past, it is just that he had more information than Alice.  
\par
Presumably, Einstein would not be happy with this state of affairs.  Quantum erasure seems to confirm that the complementarity principle is indeed a fundamental part of quantum theory.  Quantum physics has in its realm some strange consequences if one insists on using concepts from classical physics.  The founding fathers were certainly aware of this nearly a century ago.   Nowadays, physicists have learned to accept the fact that the laws of classical physics do not necessarily apply to the quantum world.  We have become much more comfortable with the ``quantum weirdness".  
\par
The quantum eraser and other experiments have done much to illustrate the dual nature of quantum theory. However, physicists today are still unable to explain why wave-particle duality exists.  In this respect, it seems that we have not come too far since the 1960's, when Richard Feynman stated, in Feynman Lectures on Physics:  ÒWe cannot make the mystery go away by explaining how it works.  We will just tell you how it works." 
Yet great progress has been made.  Understanding that it is not the uncertainty principle, but rather quantum entanglement responsible for complementarity is an enormous step, presumably in the right direction.  Quantum entanglement is at the heart of the modern theory of quantum measurement.  We have learned that it is the act of measurement itself, and not the Òquantum uncertainty" involved with the measurement that is responsible for the complementarity principle.  This may seem like a subtle point, but it is one that has caused many physicists to sleep more soundly at night.           

\begin{acknowledgments}
This research was performed in the Quantum Optics Laboratory at the Universidade Federal de Minas Gerais (UFMG) in Belo Horizonte, Minas Gerais, Brazil, with finanical support from the Brazilian funding agencies CNPq and CAPES.   
\end{acknowledgments}        

\end{document}